# The random lattice as a regularization scheme

B. Allés

*Dipartimento di Fisica dell'Università*

*Piazza Torricelli 2*

*56100-Pisa, Italy*

## Abstract

A semi-analytic method to compute the first coefficients of the renormalization group functions on a random lattice is introduced. It is used to show that the two-dimensional $O(N)$ non-linear $\sigma$-model regularized on a random lattice has the correct continuum limit. A degree $\kappa$ of "randomness" in the lattice is introduced and an estimate of the ratio $\Lambda_{random}/\Lambda_{regular}$ for two rather opposite values of $\kappa$ in the $\sigma$-model is also given. This ratio turns out to depend on $\kappa$.

# 1. Introduction

The lattice has extensively been used in numerical simulations and has proven to be a unique tool for studying the nonperturbative aspects of a field theory [1,2]. However, some of the Monte Carlo simulations have provided only a qualitative explanation or at most the order of magnitude of the quantity under study. This problem holds for example for the work done to extract expectation values of composite operators (topological susceptibility, gluon condensate, etc.). In these simulations, the physical signal is hidden by mixings whose perturbative expansions are badly known since few terms have been calculated (and the series are at best asymptotic). The random lattice [3] could be useful in this respect. Indeed, the scaling window on random lattice QCD is strongly shifted towards small values of the coupling constant [4,5]. This fact may allow a better control on the perturbative expansions which mask the nonperturbative signal (perturbative tails, mixing with other operators, etc.) as well as on the non-universal terms in the scaling function. Obviously, a careful analysis must be performed in each case. In particular, for QCD, the onset of asymptotic scaling should be deeply studied on a random lattice.

The coefficients in the perturbative expansions should not be larger on random lattices than on regular lattices, otherwise the advantage of working with smaller coupling constants is lost. One of these coefficients, the first term of the perturbative tail of the plaquette operator, can be easily computed by using the equipartition theorem. On the 4 dimensional random lattice, it is $0.1364(N^2 - 1)$ for $SU(N)$ QCD [6], while on the regular lattice it is $0.25(N^2 - 1)$. This is good news. However, an explicit calculation in each case must be done in order to decide whether the expansion has a better behaviour or not.

A question raises naturally at this point: how do we compute the coefficients of the expansions? On regular lattices, the perturbative expansions beyond 4 or 5 loops become rather awkward due to the fact that vertices develop a huge quantity of rotationally non-invariant terms [7]. If one managed to get rid of these terms, thus having vertices with the same size as in the continuum, computations would become much more feasible. The almost-rotational invariance of the random lattice could be exploited in this sense [8]. Indeed, in contrast with the regular lattice, the random lattice displays a higher degree of rotational invariance. In particular, from the nonperturbative point of view, numerical studies of nonabelian gauge theories have revealed the absence of a roughening transition [6,9] from strong to weak coupling regimes, thus opening the possibility for a wider scaling window.

Despite these potential advantages, the random lattice has been barely exploited in numerical simulations. Especially, some of its properties have been little checked. For instance, the continuum limit of models defined on random lattices has not yet been completely clarified. For statistical models with power law scaling behaviour for the correlation length, the Harris's criterion [10] predicts whether universality on a lattice with impurities holds or not if the specific heat critical exponent $\alpha$ does not vanish. For the two-dimensional Ising model (an example of vanishing $\alpha$) recent simulations [11,12] have shown that the critical behaviour depends on how one introduces the "randomness".

Asymptotically free field theories regularized on a lattice provide another important kind of models. In this case, the correlation length shows an exponential dependence on the inverse of the coupling constant. An extension of the Harris's argument [13] supports the scenario where the regular and random lattice regularizations share the same continuum limit. However, this is rather a physical consistency argument and an independent check is worthwhile.

In this work, we give an analytical proof of the universality for a field theory regularized on a random lattice. We have performed a weak coupling calculation of the first coefficients of the renormalization group functions for the two-dimensional $O(N)$ non-linear $\sigma$-model regularized on a random lattice. We have followed a semi-analytical procedure which will become clear along with the explicit calculation explained in sections 4 and 5. The action for the $O(N)$ non-linear $\sigma$-model in the continuum is

$$\mathcal{S} = \frac{1}{2g} \int d^2x (\partial_\mu \vec{\phi})^2. \qquad (1)$$

The arrow on the field $\vec{\phi}$ stands for its $N$ components which are constrained by the condition $\vec{\phi}(x)^2 = 1$ at every space-time point $x$. In Eq. (1), $g$ is the bare coupling constant. We have chosen this model because it has some properties in common with QCD. It has a rich topological structure for $N = 3$ [14], it presents spontaneous mass generation and is asymptotically free [15,16,17] in two dimensions for $N \geq 3$. This last property is the one we are interested in.

We have defined the random lattice as in reference [3]. We have also included a parameter $\kappa$ to describe the degree of "randomness" of the random lattice. This parameter is essentially the minimum distance between sites. We have checked that the same continuum limit is reached for several lattices defined with different values of $\kappa$.

The previous computations also provide the ratio between the renormalization group invariant $\Lambda$ parameters $\Lambda_{random}/\Lambda_{regular}$. This ratio is an essential ingredient if asymptotic freedom is to be checked. We have shown that this ratio depends on $\kappa$. This is an important result, mainly if, as originally proposed by T. D. Lee and collaborators and afterwards used in simulations with fermions, one is interested in averaging among different random lattices. Indeed, this $\kappa$ dependence means that one has to average only among lattices with the same $\kappa$. One can also take advantage of this dependence to improve the asymptotic scaling of the theory in the way we will discuss in section 6.

Before ending this introduction, we make some comments on the feasibility of a simulation on random lattices. Any numerical study of a theory on a random lattice is necessarily more involved than on regular lattices. Firstly, one has to store the information about links, plaquettes, etc. which can be rather memory-consuming. Another problem is the parallelization of the updating algorithm. If one can use a cluster algorithm, then the implementation is trivial and only a quite small increase in CPU-time per updating is seen (due to the larger coordination number on a random lattice in any dimensions). In QCD, instead, one usually makes use of local updating algorithms that need be parallelized to become efficient. On a regular lattice, this parallelization is well-known and can be implemented by gathering all sites in two sets, called colours. On a random lattice, a similar task can be done [18]. The number of colours on a random lattice is larger than on a regular lattice, therefore an unavoidable slowing down in the performance is detected. However, this slowing down seems not so drastic to prevent the use of random lattices.

The paper is organized as follows. In section 2 the random lattice, the parameter $\kappa$ and the action for the $\sigma$-model are introduced. This action on the random lattice is written in a convenient form for the subsequent calculation in section 3. In section 4 the general expresion for the bare 1-loop propagator on the random lattice is written. In section 5 we perform the numerical calculation of this 1-loop expression and of the corresponding renormalization constants. From them, the first coefficients of the $\beta$ and $\gamma$ functions are computed as well as the ratio between $\Lambda$ parameters. The discussion of these results and conclusions are presented in section 6.

## 2. The random lattice and the action of the $\sigma$-model

In order to construct the random lattice, we perform the following two steps. Firstly we fill the volume $V$ of the lattice with $\mathcal{N}$ sites and then a triangularization process [3] is applied to these sites.

To fill the lattice volume, we defined a minimal distance between sites by demanding that there are no two sites closer than $\sqrt{V/\mathcal{N}}/\kappa$, where $\kappa$ is some parameter to be fixed. The larger the value of $\kappa$ is, the more irregular the pattern of sites looks. Instead, for small values of $\kappa$, the sites tend to be more evenly arranged and to stay far from each other. Hence, such random lattices with small $\kappa$ can be compared with the lattices constructed in reference [19] by using the eigenvalues of complex random matrices. Moreover, the parameters defining the structure of the random lattice (i.e.: link lengths, plaquette areas, etc.) show less variance for small $\kappa$.

The smaller possible value of $\kappa$ is 1; at this value we recover the distance between sites on the regular square lattice.

Once the sites have been placed on the lattice, we proceed to construct the triangularization [3]. We followed the method of reference [9]. It consists in joining sets of three sites to form a triangle with the only condition that the circle circumscribed by these three points does not contain any other site. The sides of that triangle are the links joining the three sites and the triangle itself is a plaquette. This construction fills the whole lattice with no overlapping among the triangles. We imposed periodic boundary conditions. Thus, to construct the triangles and links, we considered the two-dimensional lattice as a torus. In figure 1 we show two $\mathcal{N} = 100$ sites random lattices, with $\kappa = 1.3$ and $\kappa = 100$.

One can define also the dual lattice. Its dual sites are the centers of the above-mentioned circles. It is clear that any link is the common side of two triangles. Thus, every link of the random lattice must be surrounded by two dual sites. The line joining these two dual sites is called the dual link. Therefore, every link is associated with a dual link.

As soon as the lattice has been constructed, one can devise tests to check the triangularization. The first and easiest one is to verify that the number of triangles (links) is equal to twice (3 times) the number of sites [3]. Other good tests are the integral properties [20]

$$\sum_j \lambda_{ij} l_{ij}^\mu = 0, \qquad \sum_{ij} \lambda_{ij} l_{ij}^\mu l_{ij}^\nu = 2V \delta^{\mu\nu}. \qquad (2)$$

In this equation, $l_{ij}^\mu$ is the link vector joining the sites $i$ and $j$ and the matrix $\lambda_{ij}$ is defined as follows:

$$\lambda_{ij} = \begin{cases} 0, & \text{if } i \text{ and } j \text{ are not linked;} \\ s_{ij}/l_{ij}, & \text{otherwise,} \end{cases} \qquad (3)$$

where $l_{ij}$ is the length of the link vector $l_{ij}^\mu$ and $s_{ij}$ is the length of the associate dual link. A third test consists in demanding that the action has only one zero mode (see next section). This last condition is very exacting and if some links lack, then spurious extra zero modes can develop.

After all of these preliminaries, we can write the action for the non-linear $\sigma$-model on a random lattice as [20]

$$\mathcal{S}^L = \frac{1}{4g} \sum_{i,j} \lambda_{ij} (\vec{\phi}_i - \vec{\phi}_j)^2. \qquad (4)$$

$g$ is the coupling constant. $i, j, \ldots$ denote sites and $\vec{\phi}_i$ means the value of the field $\vec{\phi}$ at the site $i$. The arrow on the field $\vec{\phi}$ stands for its $N$ components which, as in the continuum (see Eq. (1)), are constrained by the condition $\vec{\phi}_i^2 = 1$ at every site $i$. For a regular square lattice, $\lambda_{ij}$ is 1 for linked sites and zero otherwise. Hence, Eq.(4) becomes the standard action when it is considered on a regular lattice. One can show that the *naïve* continuum limit of Eq. (4) is the correct action for the model in the continuum, Eq. (1).

### 3. The Feynman rules on a random lattice

In this section, we will take advantage of the constraint $\vec{\phi}^2 = 1$ to rewrite the action in a more tractable form to be used in perturbation theory.

By using the constraint, the field can be written as $\vec{\phi} = (\vec{\pi}, \sqrt{1 - \vec{\pi}^2})$ where $\vec{\pi}$ is a $(N-1)$-component field. The single components of the $\vec{\pi}$ field will be denoted by capital letters $A, B, \ldots$ as in $\pi^A$. In terms of $\vec{\pi}$, the action (4) can be written as

$$\mathcal{S}^L = -\frac{1}{2g}\sum_{ij}\Lambda_{ij}(\vec{\pi}_i\vec{\pi}_j + \frac{1}{4}\vec{\pi}_i^2\vec{\pi}_j^2 + \ldots)$$
$$-\frac{1}{2}\sum_i \vec{\pi}_i^2 + \ldots, \tag{5}$$

where some constant terms have been left out. The last term comes from the jacobian of the change of variable from $\mathcal{D}\vec{\phi}$ to $\mathcal{D}\vec{\pi}$ in the functional integration. The matrix $\Lambda_{ij}$ is equal to

$$\begin{cases} \Lambda_{ij} = \lambda_{ij}, & \text{if } i \neq j; \\ \Lambda_{ii} = -\sum_j \lambda_{ij}. \end{cases} \tag{6}$$

Notice that $\sum_j \Lambda_{ij} = 0$ for all $i$.

We add an interaction with an external field $h$ to regularize the $IR$ divergences which will appear. This interaction term can be chosen as

$$\mathcal{S}^h = -\frac{h}{g}\sum_i a^2\sqrt{1-\vec{\pi}_i^2} = \frac{h}{2g}\sum_i a^2\vec{\pi}_i^2 + \frac{h}{8g}\sum_i a^2(\vec{\pi}_i^2)^2 + \ldots, \tag{7}$$

where $a$ is the lattice spacing on the random lattice defined as $a \equiv \sqrt{V/\mathcal{N}}$ and a constant has been omitted in the expansion.

Eqs. (5) plus (7) is the action we will use. It is convenient to diagonalize the matrix $\Lambda_{ij}$ which is equivalent to diagonalize the quadratic sector of the action. Let us call $-\xi_a$ the eigenvalues of $\Lambda_{ij}$ and $\mathcal{B}_{ia}$ the relevant orthonormal matrix of the change of basis. It can be proven [21] that for the matrix $\Lambda_{ij}$, all $-\xi_a$ are negative except for one of them which vanishes. We have noticed that the fact that there is only one vanishing eigenvalue is rather sensitive to the correctness of the triangularization process, (see previous section).

The index $i$ in the matrix of the change of basis $\mathcal{B}_{ia}$ stands for the space-time position (sites), while the index $a$ stands for the diagonalized basis position. Notice that this procedure is like a Fourier transformation and the diagonal basis is equivalent to the momentum space. Therefore, the $\mathcal{B}_{ia}$ matrix is equivalent to the Fourier transformation integrand $\exp(ix_ip_a)/(2\pi)$. Indices in the diagonal basis will be denoted by the letters $a,b,\ldots$ as in $\xi_a$ or $\mathcal{B}_{ia}$. Although we use the same letter, $a$, to indicate the lattice spacing and a *subindex* in the diagonal basis, there is no possible confusion. The action (Eqs. (5) and (7)) written in this basis (in momentum space) is

$$\mathcal{S}^L + \mathcal{S}^h = \frac{1}{2g} \sum_a (\xi_a + ha^2)\vec{\pi}_a^2 \qquad (8.a)$$

$$- \frac{1}{2} \sum_a \vec{\pi}_a^2 \qquad (8.b)$$

$$+ \frac{1}{8g}\{ \sum_{i,a,b,c,d} ha^2 (\vec{\pi}_a \cdot \vec{\pi}_b)(\vec{\pi}_c \cdot \vec{\pi}_d) \mathcal{B}_{ia}\mathcal{B}_{ib}\mathcal{B}_{ic}\mathcal{B}_{id}$$

$$- \sum_{i,j,a,b,c,d} \Lambda_{ij}(\vec{\pi}_a \cdot \vec{\pi}_b)(\vec{\pi}_c \cdot \vec{\pi}_d) \mathcal{B}_{ia}\mathcal{B}_{ib}\mathcal{B}_{jc}\mathcal{B}_{jd} \} \qquad (8.c)$$

$$+ \mathcal{O}(\pi^6).$$

Now, one can easily extract the Feynman rules from this lagrangian. Eq. $(8.a)$ gives the propagator

$$\Delta_a^{AB} \equiv \frac{\delta^{AB} g}{\xi_a + ha^2}, \qquad (9)$$

while Eq. $(8.b)$ is a measure term, similar to a counterterm, with an extra power of the coupling constant $g$. The four-legged vertices are in Eq. $(8.c)$. One usually performs the sum on $i$ and $j$ to get the Dirac delta indicating momentum conservation. Here, this step cannot be done. Vertices with more than four legs will not be needed in our computation.

### 4. The one loop propagator

In this section we will compute the Green function $\Gamma^{(2)}_{bare}$ at one loop. The tree order is proportional to the inverse of Eq. (9)

$$\Gamma^{(2)}_{tree\ order} = \frac{1}{g}(\frac{\xi_a}{a^2} + h). \qquad (10)$$

At one loop we have to consider the two Feynman diagrams of figure 2. The first one, figure 2.a, comes from Eq. $(8.b)$. It is the measure and gives $-1/a^2$. Eq. $(8.c)$ contributes with the tadpole shown in figure 2.b. This tadpole yields three terms, $T_1, T_2$ and $T_3$. The first one is proportional to $h$:

$$T_1 = h\left(\frac{N-1}{2} + 1\right) \sum_{i,c} \mathcal{B}_{ia}\mathcal{B}_{ib}\mathcal{B}_{ic}\mathcal{B}_{ic} \Delta_c^{AB}. \qquad (11)$$

Regularizations usually preserve the momentum conservation for every value of the cutoff. In this case, Eq. (11) turns out to be proportional to $\delta_{ab}$. This fact happens for example on a regular square lattice. On this lattice, the expresion $\Delta_{ii}^{AB} \equiv \sum_c \mathcal{B}_{ic} \Delta_c^{AB} \mathcal{B}_{ic}$ is independent of $i$ and the subsequent sum on $i$ in Eq. (11) provides the $\delta_{ab}$ because of the orthogonality of the matrix $\mathcal{B}_{ia}$. On a random lattice, we can assume that this equality is approximately true and rewrite Eq. (11) as

$$T_1 = h\left(\frac{N-1}{2} + 1\right)\delta^{AB}\Delta_0 \delta_{ab}, \tag{12}$$

where $\delta^{AB}\Delta_0$ is the $i$-independent value of $\Delta_{ii}^{AB}$. Now, we need a simple definition for this $\Delta_0$. To this purpose, let us remember that among the eigenvalues $\xi_a$ there was a zero mode. This is due to the fact that $\sum_j \Lambda_{ij} = 0$ for every $i$. Therefore, the eigenvector associated to this eigenvalue must be of the form $(1/\sqrt{\mathcal{N}}, 1/\sqrt{\mathcal{N}}, \ldots)$. Then, using the previous definition of $\Delta_{ii}^{AB}$, we can define $\Delta_0$ as

$$\Delta_0 \equiv \frac{1}{\mathcal{N}} \sum_c \frac{1}{\xi_c + ha^2}. \tag{13}$$

The validity of this assumption must be explicitly verified. For a random lattice of $40^2$ sites, Eqs. (11) and (12) differ by less than one per mille when using Eq. (13).

The terms $T_2$ and $T_3$ do not display a linear dependence on $h$. The first one is proportional to

$$T_2 \propto \sum_{i,j} \Lambda_{ij} \Delta_{jj}^{AB} \mathcal{B}_{ia}\mathcal{B}_{ib}. \tag{14}$$

Under the assumption of $j$-independence for $\Delta_{jj}^{AB}$, this term can be approximated by

$$T_2 \propto \delta^{AB}\Delta_0 \sum_i \mathcal{B}_{ia}\mathcal{B}_{ib} \sum_j \Lambda_{ij}, \tag{15}$$

and this last sum over $j$ vanishes. Hence, without the assumption, one expects that $T_2$ is negligible. Indeed, an exact calculation of the Eq. (14) on a random lattice of $40^2$ sites, shows that it is less than $1^0/_{00}$ of $T_3$. Thus, we will not consider $T_2$ any more.

The $T_3$ term is

$$T_3 = -\frac{1}{a^2} \sum_{i,j,c} \Lambda_{ij}\Delta_c^{AB}\mathcal{B}_{ia}\mathcal{B}_{jb}\mathcal{B}_{ic}\mathcal{B}_{jc}. \tag{16}$$

This equation needs be written in a more manageable form. On a regular square lattice, $\Delta_{ij}^{AB} \equiv \sum_c \mathcal{B}_{ic} \Delta_c^{AB} \mathcal{B}_{jc}$ is constant if $i$ and $j$ are linked sites. On a random lattice we can assume a similar behaviour and call $\delta^{AB}\Delta_1$ such constant. If we further assume that $\Lambda_{ii}$ can also be substituted by an $i$-independent constant, $-\bar{\lambda}$ (on our random lattice this is true within a 1%; on a regular lattice $\bar{\lambda}$ is exactly equal to 4), then Eq. (16) can be easily rewritten as

$$T_3 = \frac{1}{a^2}\delta^{AB}\delta_{ab}\Big(\Delta_1 \xi_a + \bar{\lambda}(\Delta_0 - \Delta_1)\Big). \tag{17}$$

Below, we will give explicit expressions for $\Delta_1$ and $\bar{\lambda}$. Now, let us prove Eq. (17). Splitting the sum in Eq. (16) in a term for $i \neq j$ and another for $i = j$ and using the previous definitions for $\Delta_{ij}^{AB}$ and $\Delta_{ii}^{AB}$, we get

$$T_3 = -\frac{1}{a^2}\delta^{AB}\Big(\sum_{i \neq j}\Lambda_{ij}\Delta_1 \mathcal{B}_{ia}\mathcal{B}_{jb} + \sum_i \Lambda_{ii}(\Delta_0 - \Delta_1)\mathcal{B}_{ia}\mathcal{B}_{ib} + \sum_i \Delta_1 \Lambda_{ii}\mathcal{B}_{ia}\mathcal{B}_{ib}\Big). \tag{18}$$

Now, recalling that $\Delta_1$ and $\Delta_0$ are constant, we obtain from the first and last sums $\sum_{ij}\Lambda_{ij}\mathcal{B}_{ia}\mathcal{B}_{jb} = -\xi_a \delta_{ab}$. The second term in Eq. (18) just gives $\Lambda_{ii}\delta_{ab}(\Delta_0 - \Delta_1)$. Collecting all of theses pieces, we obtain Eq. (17).

Let us write explicit definitions for both $\Delta_1$ and $\bar{\lambda}$. Recall that the eigenvector associated to the zero eigenvalue is $(1/\sqrt{\mathcal{N}}, 1/\sqrt{\mathcal{N}}, \dots)$. By using this expression in Eq. (16), we can rewrite the $\xi_a$-independent term in Eq. (17) as

$$\frac{1}{a^2}\delta^{AB}(1 - ha^2 \Delta_0). \tag{19}$$

Comparing this equation with Eq. (17) yields our definition of $\Delta_1$,

$$\Delta_1 = \Delta_0(1 + \frac{ha^2}{\bar{\lambda}}) - \frac{1}{\bar{\lambda}}. \tag{20}$$

The simplest definition for $\bar{\lambda}$ is $\bar{\lambda} = -\sum_i \Lambda_{ii}/\mathcal{N}$. However, Eq. (20) suggests this second definition: $1/\bar{\lambda} = -(\sum_i 1/\Lambda_{ii})/\mathcal{N}$. Let us call $\bar{\lambda}_1$ and $\bar{\lambda}_2$ these two possibilities. We will choose the definition which makes Eq. (17) closer to the true value, Eq. (16). As we will see in next section, we will work with two values of $\kappa$: $\kappa = 1.3$ and $\kappa = 100$. For $\kappa = 1.3$, both $\bar{\lambda}_1$ and $\bar{\lambda}_2$ coincide within errors; but for $\kappa = 100$ they yield different numbers. Using any of them, the agreement between Eq. (16) and (17) is quite good: within $\sim 5\%$, but $\bar{\lambda}_2$ shows the better agreement. Hence, we chose $\bar{\lambda}_2$ as our definition.

Moreover, an explicit numerical study on a random lattice of $40^2$ sites shows that the value of Eq. (16) for $a \neq b$ divided by the corresponding value for $a = b$ is $\sim 10^{-4}$. This favours the presence of the $\delta_{ab}$ in Eq. (17).

Collecting all pieces, the 1-loop contribution to $\Gamma^{(2)}_{bare}$ is

$$\Gamma^{(2)}_{1\ loop} = \delta^{AB}\delta_{ab}\Big(-\frac{1}{a^2} + h(\frac{N-1}{2}+1)\Delta_0 + \Delta_1\frac{\xi_a}{a^2} + \frac{1}{a^2}(1-ha^2\Delta_0)\Big). \qquad (21)$$

The procedure to know the numerical value of Eq. (21) will be the following. We will construct an explicit random lattice and calculate the matrix $\Lambda_{ij}$. The eigenvalues of this matrix will be used to compute $\Delta_0$. The diagonal of $\Lambda_{ij}$ will provide $\bar{\lambda}$ and the value of $\Delta_1$ through Eq. (20). This will be done in next section.

However, there is one more problem to be solved. To compute the renormalization constants, we are interested in the divergent terms of $\Gamma^{(2)}_{bare}$. If the behaviour of the eigenvalues $\xi_a$ near the continuum limit can be written as a power series in $p^2$, where $p$ is the momentum in the continuum, one can reasonably assume [22] that $\Delta_0$ as a function of $ha^2$, is expressed as a series:

$$\Delta_0(ha^2) = \sum_{n\geq 0}(ha^2)^n c_n + \ln(ha^2)\sum_{n\geq 0}(ha^2)^n c'_n. \qquad (22)$$

To compute the renormalization constants we are interested only in the $c_0$ and $c'_0$ terms. To extract them we will calculate $\Delta_0$ for a given random lattice and for several $h$. Then we fit Eq. (22) to these values of $\Delta_0(ha^2)$. Obviously, we need the values of $\Delta_0(ha^2)$ for infinite size lattices. We observed that the results obtained with a lattice of $40^2$ sites were essentially the same than with larger lattices if $ha^2 \geq 0.1$. That means that our finite size parameter $hV$ is of the order of $\sim 160$ or bigger. Moreover, to make the fit reliable with only the first terms of the series in Eq. (22), we cannot consider large values of $h$. As a consequence, we are forced to work in the window $ha^2 \in [0.1, 0.9]$ when doing the fits. We used least-squares fits. Alternatively, one could also make use of a linear system of equations: the values of $\Delta_0(ha^2)$ for $m$ different $ha^2$ to determine the first $m$ coefficients in Eq. (22). We obtained similar results than with the least-squares fit.

## 5. The numerical results

In this section we will explicitly construct the random lattice and make use of the definitions of $\Delta_0$ and $\Delta_1$ in Eq. (13) and (20) to know the numerical value of Eq. (21).

We used random lattices of $40^2$ sites. This lattice size is enough to obtain reasonable results as discussed in the previous section. We chose two different $\kappa$ parameters: $\kappa = 1.3$ and $\kappa = 100$. Then, for each value of this parameter, we constructed 40 random lattices and determined their respective matrix $\Lambda_{ij}$. These matrices were diagonalized to compute $\Delta_0$ by means of Eq. (13) for the following 16 values of $ha^2$: $\{0.1, 0.15, 0.2, 0.25, 0.3, 0.35, 0.4, 0.45, 0.5, 0.55, 0.6, 0.65, 0.7, 0.75, 0.8, 0.85\}$. The value of $\bar{\lambda}$ was also extracted. The diagonalization of the matrix is performed with a NAG subroutine and it takes $\sim 23$ CPU seconds for our lattice size with a CRAY C-94/2128.

At this point, we have 40 independent determinations of $\Delta_0$ for each value of $ha^2$. Then, the final result for $\Delta_0(ha^2)$, to be used in the fit, is extracted simply by computing the mean and statistical error from this set of 40 determinations. We noticed that there is no correlation among these 40 values, so we assumed a normal distribution to compute the errors.

These 16 values of $\Delta_0(ha^2)$ both for $\kappa = 1.3$ and $\kappa = 100$ are shown in table 1. From this table it is clear that the regularization depends only on $\kappa$. This statement is supported by the fact that the determination of $\Delta_0$ is rather insensitive to the particular random lattice used, as the errors indicate.

Now, the 16 values of $\Delta_0(ha^2)$ are fitted to the first coefficients in the series of Eq. (22). To accept the result of the fit, we demanded that the values obtained for these coefficients remain stable when we remove or add some of the 16 fitted points. The value of the $\chi^2$ is not a good test of reliability for the fit because we are not fitting measured quantities with a gaussian distribution of errors. Indeed, we always obtained a very small $\chi^2$. The best result was obtained when fitting five coefficients in Eq. (22). Here we report all of these five coefficients, but recall that only the first two ($c_0$ and $c'_0$) are interesting for us:

$$\Delta_0(ha^2) = 0.2825(29) - 0.07964(83)\ln(ha^2) - 0.0255(30)\, ha^2$$
$$+ 0.0141(35)\, ha^2 \ln(ha^2) - 0.0007(20)\, (ha^2)^2 \ln(ha^2) + \mathcal{O}((ha^2)^2) \quad (23)$$
$$\bar{\lambda} = 3.887(7),$$

for $\kappa = 1.3$ and

$$\Delta_0(ha^2) = 0.246(11) - 0.0822(30)\ln(ha^2) - 0.017(11)\, ha^2$$
$$+ 0.017(13)\, ha^2 \ln(ha^2) - 0.0035(75)\, (ha^2)^2 \ln(ha^2) + \mathcal{O}((ha^2)^2) \quad (24)$$
$$\bar{\lambda} = 5.42(6),$$

for $\kappa = 100$. Here the errors are larger because of the greater irregularity among the 40 random lattices. The goodness of the fit is manifested by the fact that the next-to-leading coefficients, ($c_i$ and $c'_i$ for $i \geq 1$) are clearly decreasing for both $\kappa = 1.3$ and $\kappa = 100$. The 16 values of $\Delta_0(ha^2)$ together with the results of the fit, Eqs. (23) and (24), are shown in figure 3 for both values of $\kappa$. We point out again the manifest difference between the two cases.

However, notice that the coefficients proportional to the divergence in Eqs. (23) and (24) are compatible within errors. This is welcome because this coefficient is proportional to the first terms of the beta and gamma functions, $\beta_0$ and $\gamma_0$. For a regular square lattice this coefficient is [16,17,23] $-1/(4\pi) = -0.07957$, in good agreement with the values showed in Eqs. (23) and (24). This is the numerical result which supports the fact that the two-dimensional $O(N)$ nonlinear $\sigma$-model regularized on a random lattice shares the same universality class than regularized on a regular square lattice.

Now, using Eqs. (20), (23) and (24) into Eq. (21), one obtains the bare Green function. We can renormalize it in the $\overline{\text{MS}}$-scheme

$$\Gamma_r^{(2)\overline{\text{MS}}}(p^2, g_r, h_r, \mu) = Z_\pi \Gamma_{bare}^{(2)}(\xi, g, h, a)\Big|_{\substack{g = g_r Z_g \\ h = h_r Z_g Z_\pi^{-1/2} \\ \xi = a^2 p^2}}, \quad (25)$$

where $\Gamma_r^{(2)\overline{\text{MS}}}$ is

$$\Gamma_r^{(2)\overline{\text{MS}}} = \frac{p^2 + h_r}{g_r} + (p^2 + \frac{N-1}{2} h_r)\frac{1}{4\pi}\ln(\frac{\mu^2}{h_r}). \quad (26)$$

Writing the renormalization constants as $Z_g \equiv 1 + z_g g$ and $Z_\pi \equiv 1 + z_\pi g$, we obtain

$$z_g = (N-2)\Big(\frac{1}{4\pi}\ln(\frac{\mu^2}{h_r}) - c_0 - c_0'\ln(h_r a^2)\Big) - \frac{1}{\bar\lambda},$$
$$z_\pi = (N-1)\Big(\frac{1}{4\pi}\ln(\frac{\mu^2}{h_r}) - c_0 - c_0'\ln(h_r a^2)\Big). \qquad (27)$$

Now, using the definition $\beta^L \equiv -a\,d/da\ g = -\beta_0 g^2 - \beta_1 g^3 - \ldots$ for the beta function and $\gamma^L \equiv a\,d/da\ \ln Z_\pi = \gamma_0 g + \gamma_1 g^2 + \ldots$ for the gamma function, we get

$$\beta_0 = -2c_0'(N-2), \qquad \gamma_0 = -2c_0'(N-1). \qquad (28)$$

Finally, using well known methods [24,25], we obtain the ratio between renormalization group invariant mass parameters, $\Lambda_{random}/\Lambda_{regular}$. To perform this computation, we used the exact values of $c_0 = 5\ln 2/(4\pi)$, $c_0' = -1/(4\pi)$ and $\bar\lambda = 4$ for a regular square lattice and we assumed the same value of $c_0'$ for random lattices as well in order to obtain finite results for the ratio. This ratio is

$$\kappa = 1.3 \qquad \Lambda_{random}/\Lambda_{regular} = \exp\Big(-0.042(18) - \frac{0.046(3)}{N-2}\Big),$$
$$\kappa = 100. \qquad \Lambda_{random}/\Lambda_{regular} = \exp\Big(0.19(7) + \frac{0.41(1)}{N-2}\Big). \qquad (29)$$

For the $O(3)$ non-linear $\sigma$-model, these expressions become 0.92(2) and 1.8(2) respectively. These numbers differ each other more than four standard deviations. Had we used the definition $\bar\lambda_1$, then similar values would have been obtained: 0.92(2) and 2.1(2) respectively. The ratios we have got can be compared with those of ref. [26] obtained with simulations which give $\Lambda_{random}/\Lambda_{regular} \approx 1.29$; (in this reference the correlation length is measured on rather small lattice and no finite size analysis is done).

The fact that by varying the $\kappa$ parameter, one obtains different regularizations of the same theory opens some prospects to improve the efficiency of simulations. They will be examined in the next section.

## 6. Discussion and conclusions

A family of random lattices has been proposed as a regularization scheme for a field theory. This family is characterized by a parameter $\kappa$ that gives a measure of the "randomness". We have checked that the correct continuum limit is reached for a two-dimensional $O(N)$ non-linear $\sigma$-model for two quite different values of $\kappa$. The correctness of this statement is measured by the coefficients of the logarithm in Eqs. (23) and (24) which should be equal to $-0.07957$. Indeed, the first universal terms of the $\beta$ and $\gamma$ functions are proportional to this coefficient.

We have also evaluated the ratio between lambda parameters for a random lattice and a regular square lattice. We have discovered that this ratio depends on $\kappa$. This result means that lattices with different $\kappa$ are different regularizations. Hence, if an average among several random lattices is to be done, one must choose a fixed value of $\kappa$ for all of them. On the other hand, this dependence on $\kappa$ can also have useful consequences. Let us begin by discussing them when $\Lambda_{random} > \Lambda_{regular}$. In this case, the scaling window shifts towards larger values of the bare coupling $g$. Recalling that the roughening transition is absent on random lattices (for theories where it has no physical meaning) [6,9], one can imagine a scenario where strong coupling expansions become useful in this scaling window. If single observables are analysed, all conclusions are subject to the fulfilment of the asymptotic scaling condition.

Let us analyse the case $\Lambda_{regular} > \Lambda_{random}$. If this happens, then the scaling window is placed in a region of lower values of $g$. This location could favour the onset of asymptotic scaling, although a rigorous numerical study is needed in each case to control the non-universal terms in the $\beta$ function. Working with lower coupling constant may also simplify the subtraction of perturbative expansions from the Monte Carlo signal.

Actually, the ratio $\Lambda_{random}/\Lambda_{regular}$ for the $O(N)$ non-linear $\sigma$-model in two dimensions has proven to be quite close to 1. Thus, the previous considerations are of rather marginal interest for this model. However, in QCD they could be much more relevant. Indeed, previous simulations [4,5] have shown that the ratio of lambda parameters is very small: $\mathcal{O}(10^{-1})$ and $\mathcal{O}(10^{-2})$ for $SU(2)$ and $SU(3)$ QCD respectively. Hence, the $\kappa$-dependence of this ratio can be more dramatic. In particular, for these values of the ratio, the scaling window is strongly shifted towards the region of small $g$, opening the second scenario outlined above.

The fact that the ratio of lambda parameters is much less than 1 in QCD can be easily understood by geometrical arguments [27]. The size of a lattice is usually given in units of sites. However, this has no much sense in QCD because for a gauge theory, the fields live on the links. So, one should give the size of a lattice as the number of links. On a random lattice in four dimensions there are four times more links than on the corresponding regular square lattice with the same number of sites. This explains why the lattice spacing on a random lattice is longer than on a regular square lattice. For a scalar theory, the degrees of freedom live on the sites and the previous argument does not work.

A numerical study of the consequences of the $\kappa$-dependence in QCD is in progress.

**Acknowledgements**

I thank Matteo Beccaria, Adriano Di Giacomo, Domènec Espriu, Pedro Garrido, Miguel Angel Muñoz and Haris Panagopoulos for useful conversations.

**Figure Captions**

Figure 1. Two random lattices with $10^2$ sites. The lattice $a)$ was obtained with $\kappa = 1.3$ and the $b)$ with $\kappa = 100$. For both of them we used the random number generator RANF and the triangularization method of ref. [3,9].

Figure 2. Feynman diagrams contributing to the 1 loop propagator. The cross stands for the measure vertex.

Figure 3. $\Delta_0$ versus $ha^2$ for $\kappa = 1.3$ and $\kappa = 100$. The crosses represent the numerical results of table 1. The solid and dashed lines are the curves derived from the fits, Eqs. (23) and (24) respectively.

**Table Caption**

Table 1. Values of $\Delta_0(ha^2)$ for $\kappa = 1.3$ and $100$.

**Table 1**

| $ha^2$ | $\Delta_0(ha^2, \kappa = 1.3)$ | $\Delta_0(ha^2, \kappa = 100)$ |
|---|---|---|
| 0.10 | 0.46006(4) | 0.4301(1) |
| 0.15 | 0.42573(4) | 0.3950(1) |
| 0.20 | 0.40103(4) | 0.3700(1) |
| 0.25 | 0.38165(4) | 0.3504(1) |
| 0.30 | 0.36566(4) | 0.3344(1) |
| 0.35 | 0.35203(4) | 0.3208(1) |
| 0.40 | 0.34015(4) | 0.3091(1) |
| 0.45 | 0.32960(4) | 0.2987(1) |
| 0.50 | 0.32012(3) | 0.2895(1) |
| 0.55 | 0.31151(3) | 0.2811(1) |
| 0.60 | 0.30362(3) | 0.2735(1) |
| 0.65 | 0.29634(3) | 0.2665(1) |
| 0.70 | 0.28958(3) | 0.2600(1) |
| 0.75 | 0.28328(3) | 0.2540(1) |
| 0.80 | 0.27738(3) | 0.2484(1) |
| 0.85 | 0.27183(3) | 0.2432(1) |

# Figure 1

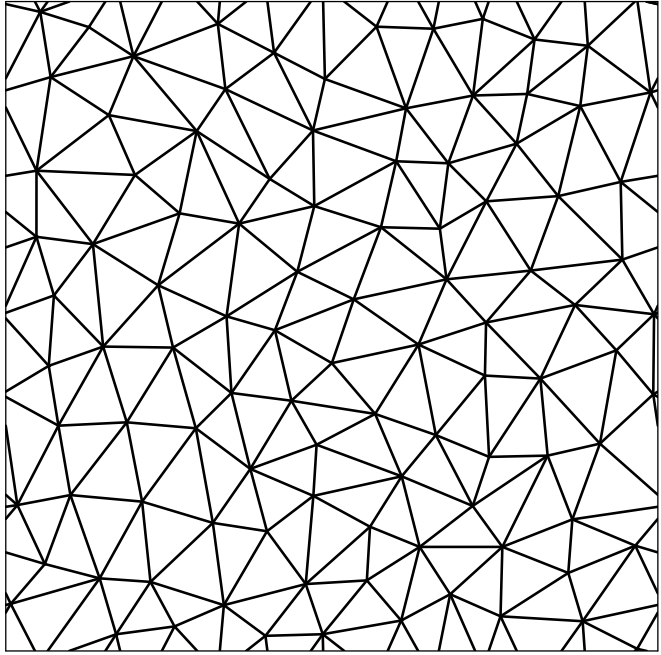

**a)**

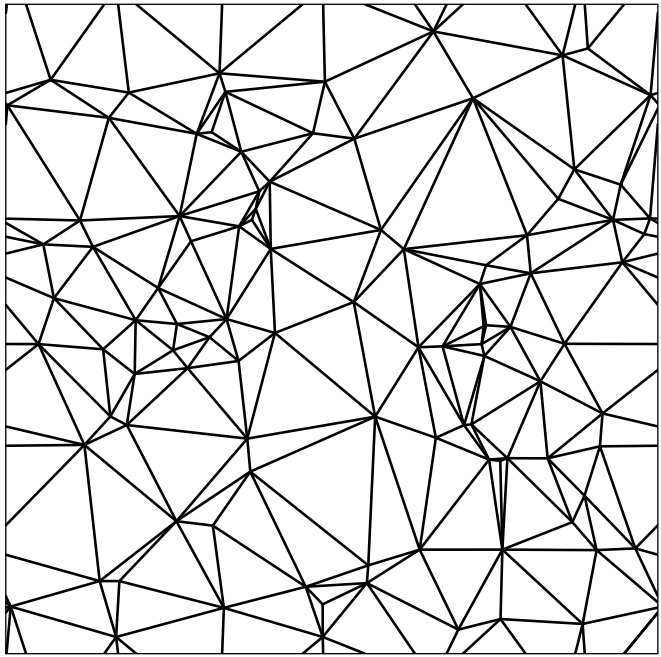

**b)**

# Figure 2

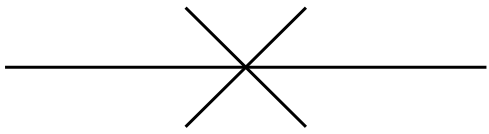

a)

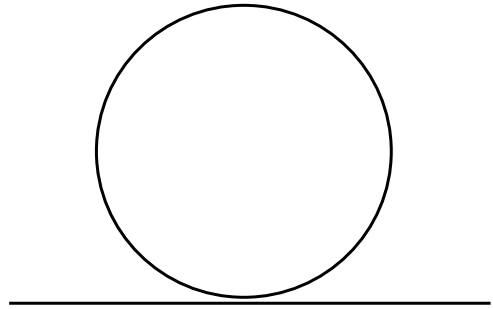

b)

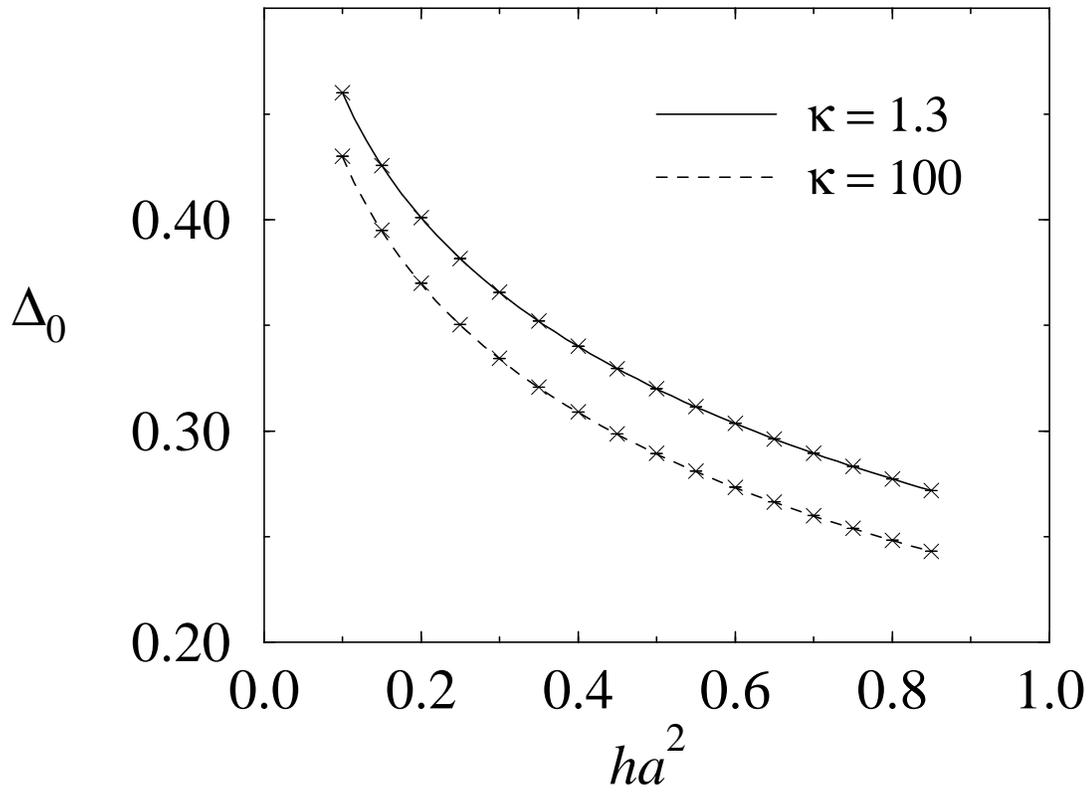